\documentclass{ws-procs10x7}

\newcommand{\BF}{\ensuremath{{\cal{B}}}}

\newcommand{\mass}{MeV/$c^2$}

\newcommand{\bckpp}{\ensuremath{B^+\to K^+\pi^+\pi^-}}
\newcommand{\bckkk}{\ensuremath{B^+\to K^+K^+K^-}}

\newcommand{\kpp}{\ensuremath{K^+\pi^+\pi^-}}
\newcommand{\kkk}{\ensuremath{K^+K^+K^-}}

\newcommand{\pipi}{\ensuremath{\pi^+\pi^-}}
\newcommand{\kpkm}{\ensuremath{K^+K^-}}

\newcommand{\chic}{\ensuremath{\chi_{c0}}}

\def\plb#1#2#3{{ Phys.\ Lett.}  {\bf B#1}, #3 (#2)}

\def\prd#1#2#3{{ Phys.\ Rev.}   {\bf D#1}, #3 (#2)}

\def\prl#1#2#3{{ Phys.\ Rev.\ Lett.} {\bf #1}, #3 (#2)}

\begin{document}

\title{DALITZ ANALYSIS OF $\bckpp$ and $\bckkk$ }

\author{Alex Bondar}

\address{(On behalf of the Belle Collaboration)\\
Budker Institute of Nuclear Physics, Lavrentieva 11, Novosibirsk
630090, Russia\\E-mail:bondar@inp.nsk.su}

\twocolumn[\maketitle\abstract{
We report results of the Dalitz analysis of the three-body charmless $\bckpp$ and
$\bckkk$ decays based on a $140$~fb$^{-1}$ data sample collected with the
Belle detector. Measurements of branching fractions for quasi-two-body
decays to scalar-pseudoscalar states: $B^+\to f_0(980)K^+$,
$B^+\to K^*_0(1430)^0\pi^+$, and to vector-pseudoscalar states:
$B^+\to K^*(892)^0\pi^+$, $B^+\to\rho^0K^+$, $B^+\to\phi K^+$ are
presented. Upper limits on decays to some pseudoscalar-tensor final
states are reported. We also report the new measurement of the $B^+\to\chic K^+$
branching fraction in two $\chic$ decay channels: $\chic\to \pi^+\pi^-$ and
$\chic\to K^+K^-$.
}] 

\section{Introduction}%1
Studies of $B$ meson decays to three-body charmless hadronic final states are
a natural extension of studies of decays to two-body charmless final states.
Some of the final states considered so far as two-body (for example, $\rho \pi$,
$K^*\pi$, etc.) proceed via quasi-two-body processes involving a wide
resonance state that immediately decays in the simplest case to two particles,
thereby producing a three-body final state. $B$ meson decays to
three-body charmless hadronic final states may
provide new possibilities for CP violation searches. 

%For example, a new method
%to extract the weak angle $\phi_3$ from the isospin analysis and measure the
%time dependent CP asymmetry in the decay $B^0\to K_S^0\pi^+\pi^-$ was suggested
%in Ref.\cite{b2hhhcp}. The measurement of angle $\phi_1$ in three-body
%$B^0\to K_S^0K^+K^-$ decay was suggested in Ref.~\cite{garmash2}.

Observation of $B$ meson decays to various three-body charmless hadronic final
states has already been reported by the Belle~\cite{garmash,chang,garmash2},
CLEO~\cite{eckhart} and BaBar~\cite{aubert} Collaborations. First results on
the distribution of signal events over the Dalitz plot in the three-body
$\bckpp$ and $\bckkk$ decays are described in Ref.~\cite{garmash}. With a data
sample of $29.1$~fb$^{-1}$ a simplified analysis technique was used
because of lack of
statistics. Using a similar technique, the BaBar collaboration has
reported results of their analysis of the Dalitz plot for the decay $\bckpp$
with a $56.4$~fb$^{-1}$ data sample~\cite{babar-dalitz}. With the large data
sample that is now available, we can perform a full amplitude analysis. The
analysis described in this paper is based on a 140\,fb$^{-1}$ data sample
containing 152 million $B\bar{B}$ pairs, collected  with the Belle detector
operating at the KEKB asymmetric-energy $e^+e^-$ collider. 

\section{Amplitude Analysis}
\label{sec:aa}
\begin{figure*}[th]
  \centering
  \includegraphics[width=0.32\textwidth]{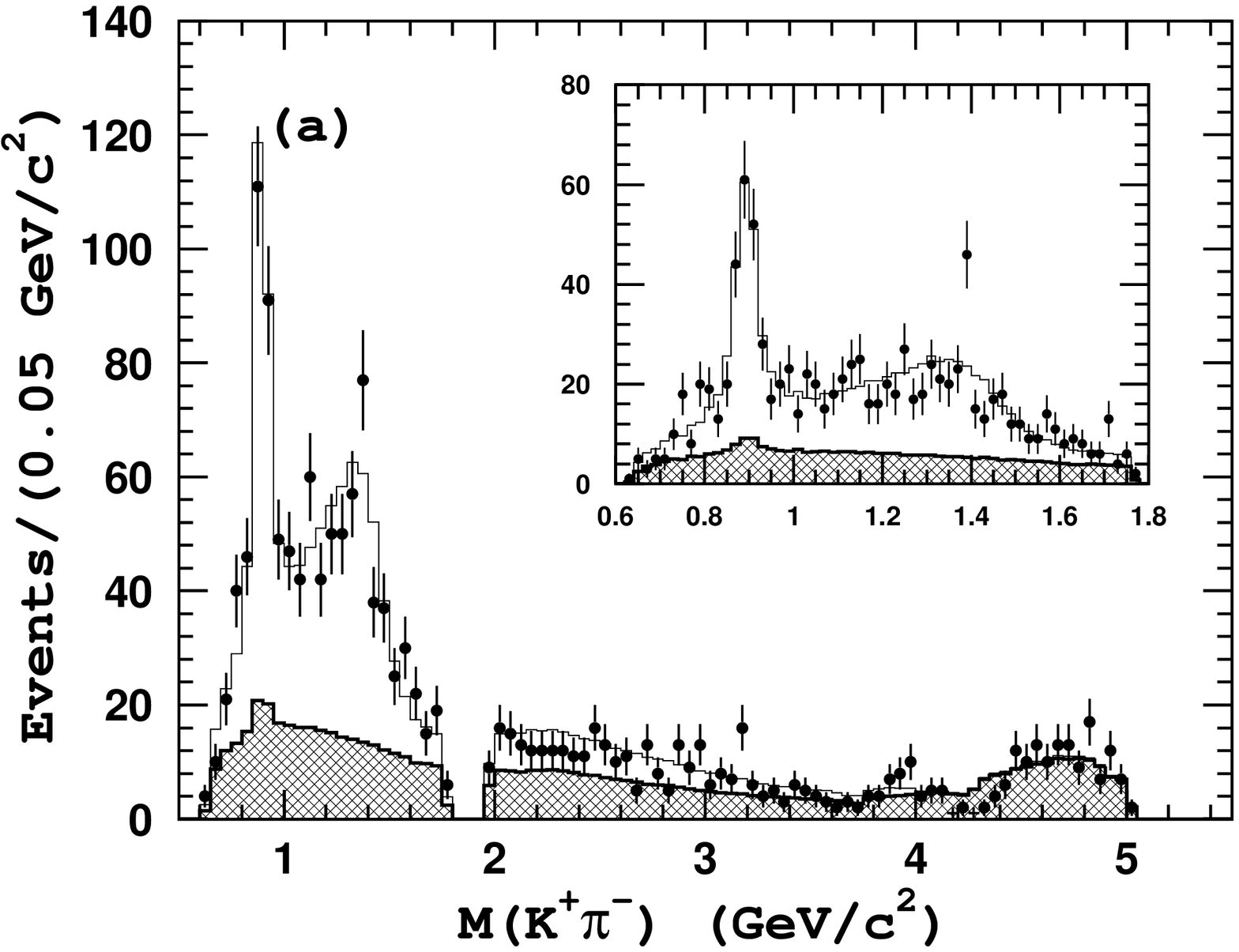}
  \includegraphics[width=0.32\textwidth]{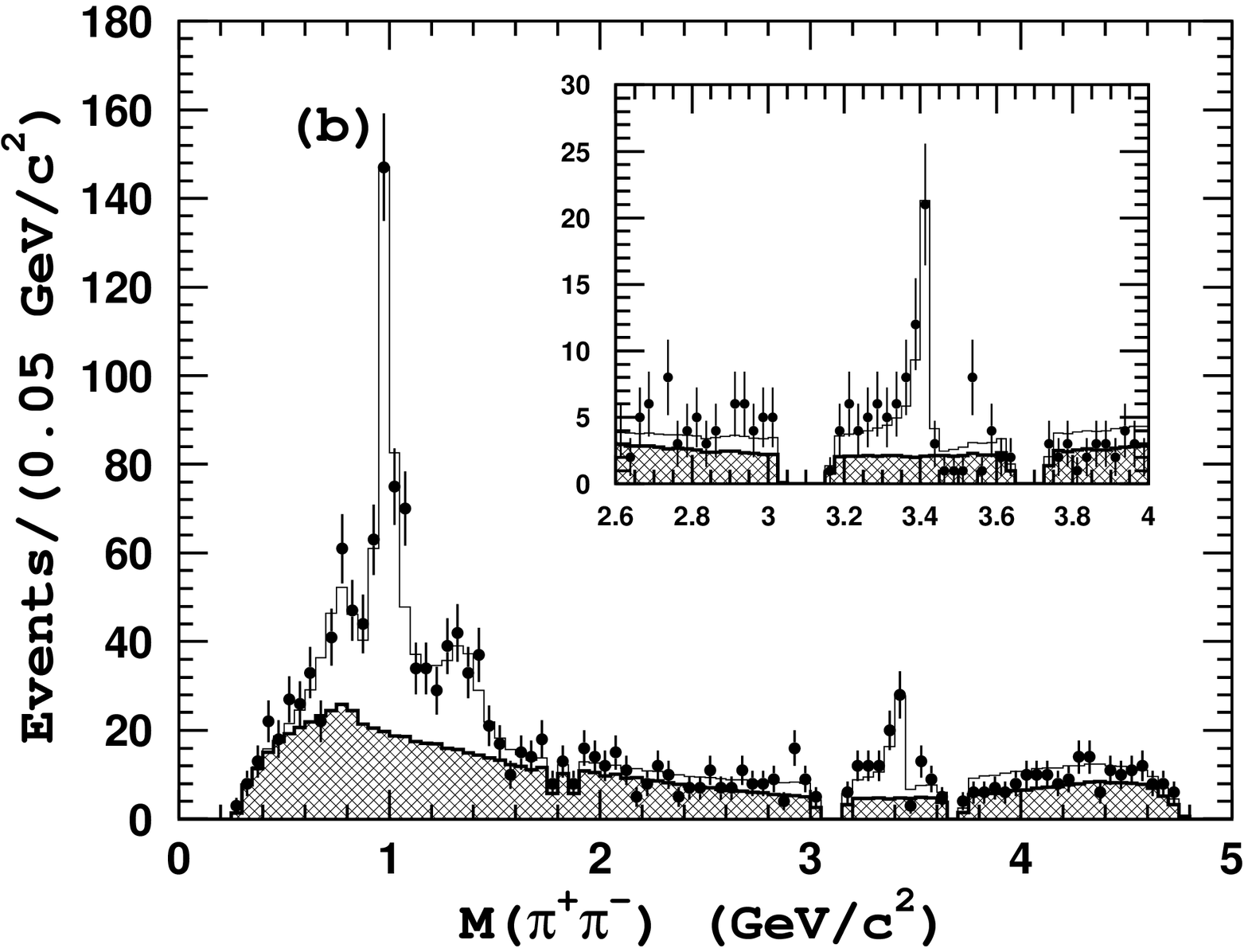}
  \includegraphics[width=0.32\textwidth]{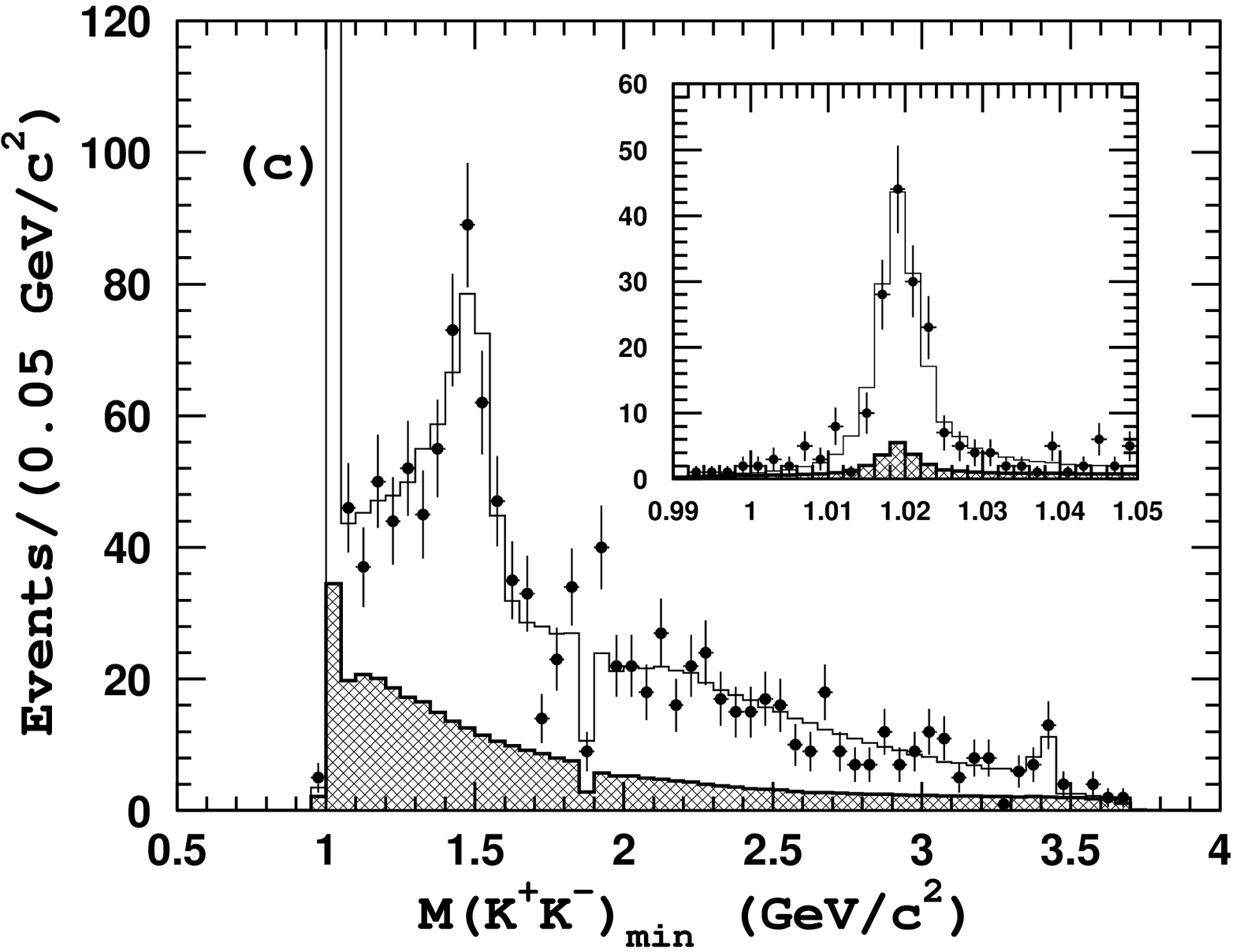} \\
  \caption{Results of the fit to $\kpp$ and $\kkk$ events. Points with
           error bars are  data, the open histograms
           are the fit result and hatched histograms are the background
           components. Insets show: (a) the $K^*(892)-K_0^*(1430)$
           mass region in 20~\mass~ bins;  (b) the
           $\chic$ mass region in 25~\mass~ bins;  (d) the
           $\phi(1020)$ mass region in 2~\mass~ bins.}
\label{fig:kpp-mods}
\end{figure*}

Analysis of two-body mass spectra shows that a
significant fraction of the signals
observed in $\bckpp$ and $\bckkk$ decays can be assigned to quasi-two-body
intermediate states. These resonances will cause a non-uniform  distribution
of events in phase space that can be analyzed using the technique pioneered by
Dalitz. Multiple resonances that occur nearby in phase space will
interfere and provide an opportunity to measure both the amplitudes and
relative phases of the intermediate states. This in turn allows us to deduce
their relative branching fractions. Details of the event selection
and amplitude analysis could be found in the Belle contributed
paper~\cite{Belle_cont}. Here we can present the main
results only. The examples of the two-body invariant mass
distributions and their description by a fit of $\bckpp$ and
$\bckkk$ decays are presented in Fig.~\ref{fig:kpp-mods}. Results on
the branching fractions for quasi-two-body decays are summarized
in Table~\ref{tab:branch}.

\begin{table*}[!ht]
  \caption{Summary of branching fraction results. The first quoted error is
           statistical, the second is systematic and the third is the model
           error.  The
           charmless total fractions in this table exclude the $\chic$
           contribution. The value given in brackets for the $K^*_0(1430)\pi^+$
           and $\chic K^+$
           channels corresponds to the second solution 
           (see~$^7$ for details).}
  \medskip
  \label{tab:branch}
\centering
  \begin{tabular}{lcr} \hline \hline
\hspace*{2mm}Mode\hspace*{3mm} &
\hspace*{0mm}$\BF(B^+\to Rh^+)$ &
\hspace*{1mm}$\BF(B^+\to Rh^+)\times10^{6}$  \\
&
\hspace*{0mm}$\times\BF(R\to h^+h^-)\times10^{6}$ &
  \\ \hline \hline
 $\kpp$ charmless total & $-$
                        & $46.6\pm2.1\pm4.3$  \\
 $K^*(892)^0\pi^+$, $K^*(892)^0\to K^+\pi^-$
                        & $6.55\pm0.60\pm0.60^{+0.38}_{-0.57}$
                        & $9.83\pm0.90\pm0.90^{+0.57}_{-0.86}$     \\
 $K^*_0(1430)\pi^+$, $K^*_0(1430)\to K^+\pi^-$
                        & $27.9\pm1.8\pm2.6^{+8.5}_{-5.4}$
                        & $45.0\pm2.9\pm6.2^{+13.7}_{-~8.7}$       \\

                        & ($5.12\pm1.36\pm0.49^{+1.91}_{-0.51}$)
                        & ($8.26\pm2.20\pm1.19^{+3.08}_{-0.82}$)   \\
 $K^*(1410)\pi^+$, $K^*(1410)\to K^+\pi^-$
                        & $<2.0$ & $-$                             \\
 $K^*(1680)\pi^+$, $K^*(1680)\to K^+\pi^-$
                        & $<3.1$ & $-$                             \\
 $K^*_2(1430)\pi^+$, $K^*_2(1430)\to K^+\pi^-$
                        & $<2.3$ & $-$                             \\
 $\rho^0(770)K^+$, $\rho^0(770)\to\pi^+\pi^-$
                        & $4.78\pm0.75\pm0.44^{+0.91}_{-0.87}$
                        & $4.78\pm0.75\pm0.44^{+0.91}_{-0.87}$     \\
 $f_0(980)K^+$, $f_0(980)\to\pi^+\pi^-$
                        & $7.55\pm1.24\pm0.69^{+1.48}_{-0.96}$
                        & $-$                                      \\
 $f_2(1270)K^+$, $f_2(1270)\to\pi^+\pi^-$
                        & $<1.3$ & $-$                             \\
 Non-resonant
                        & $-$
                        & $17.3\pm1.7\pm1.6^{+17.1}_{-7.8}$        \\
\hline
 $\kkk$ charmless total & $-$
                        & $30.6\pm1.2\pm2.3$                       \\
 $\phi K^+$, $\phi\to K^+K^-$
                        & $4.72\pm0.45\pm0.35^{+0.39}_{-0.22}$
                        & $9.60\pm0.92\pm0.71^{+0.78}_{-0.46}$     \\
 $\phi(1680)K^+$, $\phi(1680)\to K^+K^-$
                        & $<0.8$                                   \\
 $f_0(980)K^+$, $f_0(980)\to K^+K^-$                
                        & $<2.9$                                   \\
 $f'_2(1525)K^+$, $f'_2(1525)\to K^+K^-$
                        & $<2.1$                                   \\
 $a_2(1320)K^+$, $a_2(1320)\to K^+K^-$ 
                        & $<1.1$                                   \\
 Non-resonant
                        & $-$
                        & $24.0\pm1.5\pm1.8^{+1.9}_{-5.7}$         \\
\hline
 $\chic K^+$, $\chic\to\pi^+\pi^-$
                        & $1.37\pm0.28\pm0.12^{+0.34}_{-0.35}$
                        & $-$                                      \\
%                        & $2.74\pm0.56\pm0.46^{+0.68}_{-0.70}$     \\
 $\chic K^+$, $\chic\to K^+K^-$
                        & $0.86\pm0.26\pm0.06^{+0.20}_{-0.05}$
                        & $-$                                      \\
%                        & $1.45\pm0.45\pm0.25^{+0.33}_{-0.08}$     \\

                        & ($2.58\pm0.43\pm0.19^{+0.20}_{-0.05}$)
                        & $-$                                      \\
%                        & $4.35\pm0.74\pm0.74^{+0.22}_{-0.27}$     \\
 $\chic K^+$ combined   & $-$
                        & $196\pm35\pm33^{+197}_{-26}$     \\

\hline \hline
  \end{tabular}
\end{table*}

\section{Discussion}
\label{sec:discussion}

  With a 140~fb$^{-1}$ data sample collected with the Belle detector, we
have performed the first amplitude analysis~\footnote {At this
  Conference Babar presented the Dalits plot analysis for $\kpp$~\cite{Babar_kpp}.} of
$B$ meson decays to the three-body charmless $\kpp$ and $\kkk$ final
states. Such analysis often suffers from uncertainties related to the non-unique
description of the decay amplitude. In our case the uncertainty
comes mainly from the parametrization of the non-resonant amplitude.
%We use an empirical parametrization for the  
%data. In some cases the uncertainty in the description of the
%non-resonant amplitude significantly affects the extraction of relative
%fractions of other quasi-two-body channels. Further theoretical progress
%on might allow reduction of this uncertainty.
It is worth noting that fractions of the non-resonant decay in both
$\bckpp$ and $\bckkk$ decays are comparable in size and comprise a
significant fraction of the total three-body signal, which may indicate the
common nature of the amplitudes.

Despite the large model uncertainty discussed above, there is a set of
quasi-two-body signals whose branching fractions can be measured with a
relatively small model error. In particular, clear signals are observed in
the $B^+\to K^*(892)^0\pi^+$, $B^+\to\rho^0(770)K^+$, $B^+\to f_0(980)K^+$
and $B^+\to\phi K^+$ decay channels. The model uncertainty for these channels
is small due to the narrow width of the resonances and in vector-pseudoscalar
decays due to the clear signature of the vector meson polarization.

The branching fraction value measured for the decay $B^+\to K^*(892)^0\pi^+$
is significantly lower than that reported earlier~\cite{garmash,babar-dalitz}.
The simplified technique used for Dalitz analysis of the $\bckpp$ decay
described in~\cite{garmash,babar-dalitz} has no sensitivity to the relative
phases between different resonances, resulting in a large model error.
The full amplitude analysis presented in this paper consistently treats
effects of interference between different quasi-two-body amplitudes thus
reducing the model error.
The analysis suggests the presence of a large \mbox{non-$K^*(892)^0\pi^+$}
(presumably non-resonant) amplitude in the mass region of the $K^*(892)^0$ that
absorbs a significant fraction of the $B$ signal. 
The $B^+\to K^*(892)^0\pi^+$ branching fraction measured in our analysis is in
better agreement with theoretical predictions based on the QCD factorization
approach.

The decay mode $B^+\to f_0(980)K^+$ is the first observed example of a $B$
decay to a charmless scalar-pseudoscalar final state. The mass 
$M(f_0(980))=976\pm4^{+2}_{-3}$~\mass ~and width
$\Gamma(f_0(980))=61\pm9^{+14}_{-8}$~\mass ~obtained from the fit are in
agreement with previous measurements. 
The sensitivity to the $B^+\to f_0(980)K^+$ decay in the $\kkk$
final state is greatly reduced by the large $B^+\to\phi K^+$ signal and by
the scalar non-resonant amplitude. No statistically significant
contribution from this channel to the $\kkk$ three-body final state is
observed, thus only a 90\% confidence level upper limit for the
product of the  corresponding branching fractions is reported.

We report the first observation of the decay
$B^+\to\rho^0(770)K^+$. This is one of the 
channels where large direct CP violation is expected.

Due to the very narrow width of the $\phi$ meson, the branching fraction for
the decay $B^+\to\phi K^+$ is determined with a small model uncertainty.
%The obtained value is in good agreement with previous measurements~\cite{phik}.

A clear signal is also observed for the decay $B^+\to \chic K^+$ in both
$\chic\to \pi^+\pi^-$ and $\chic\to K^+K^-$ channels.
Although quite significant statistically, the $B^+\to\chic K^+$ signal
constitutes only a small fraction of the total three-body signal and thus
suffers from a large model error, especially in the $\kkk$ final state. For
this decay mode, the charmless non-resonant amplitude in the $\chic$ mass
region is enhanced compared to the $\kpp$ final state due to the interference
caused by the presence of the two identical kaons.

For other quasi-two-body channels the interpretation of fit results is less
certain. Although the $B^+\to K^*_0(1430)\pi^+$ signal is observed with a high
statistical significance, its branching fraction is determined with a large
model error. Two solutions with significantly different fractions of the
$B^+\to K^*_0(1430)\pi^+$ signal but similar likelihood values are obtained
from the fit to $\kpp$ events. A study with MC simulation confirms the
presence of the second solution. This
may indicate that in order to choose a unique solution additional external
information is required. In this sense, the most useful piece of information
seems to be the phenomenological estimation of the $B^+\to K^*_0(1430)\pi^+$
branching fraction. The analysis of $B$ meson decays to scalar-pseudoscalar
final states
described in Ref.~\cite{b2ps} suggests that the branching fraction for the
$B^+\to K^*_0(1430)\pi^+$ decay can be as large as $40\times 10^{-6}$.
Unfortunately, the predicted value suffers from a large systematic error
that is mainly due to uncertainty in the $K^*_0(1430)$ decay constant
$f_{K^*_0(1430)}$. Different techniques used to estimate
$f_{K^*_0(1430)}$~\cite{b2ps,maltman} give significantly different results.
Further improvement in this field would be useful.

We also check possible contributions from $B$ decays to various
pseudoscalar-tensor ($PT$) states.
In the factorization approximation, charmless $B$ decays
to $PT$ final states are expected to occur at
the level of $10^{-7}$ or less.  
We find no statistically significant signal in any of these
channels. As a result, we set 90\% confidence level upper limits for their
branching fractions.

We cannot identify unambiguously the broad structures observed in the
$M(\pipi)\simeq1.3$~GeV/$c^2$ mass region in the $\kpp$ final state
denoted as $f_X(1300)$ in our analysis and at $M(\kpkm)\simeq1.5$~GeV/$c^2$
in the $\kkk$ final state denoted as $f_X(1500)$. If approximated by a
single resonant state, $f_X(1300)$ is equally well described by a
scalar or vector amplitude. Analysis with higher statistics might allow
a more definite conclusion. The best description of the $f_X(1500)$
is achieved with a scalar amplitude with mass and width from the fit
consistent with $f_0(1500)$ states. 

Results of the $\bckkk$ Dalitz analysis can be also useful in connection
with the measurement of CP violation in $B^0\to K^0_SK^+K^-$ decay reported
recently by the Belle~\cite{kskk-cp-belle} and BaBar~\cite{kskk-cp-babar}
collaborations. An isospin analysis of $B$ decays to
three-kaon final states suggests the dominance of the CP-even component in the
$B^0\to K^0_SK^+K^-$ decay (after the $B^0\to\phi K^0_S$ signal is excluded).
This conclusion can be checked independently by an amplitude analysis of the
$K^0_SK^+K^-$ final state, where the fraction of
CP-odd states can be obtained as a fraction of states with odd orbital momenta.
Unfortunately, such an analysis is not feasible with the current experimental
data set. Nevertheless, the fact that we do not observe any vector amplitude
other than $B^+\to\phi K^+$ in the decay $\bckkk$ confirms the conclusion.


\begin{thebibliography}{99} 

%%%%%%%%%%%%%%%%%%%%%%%%%%%%%%%%%%%%%%%%%%%%%%%%%%%%%%%%%%%%%%%%%%%%%%%%%%%%%%%
%%%%%%%%%%%%%%%%%%%%%%%%%%%%%%  CHAPTER 1  %%%%%%%%%%%%%%%%%%%%%%%%%%%%%%%%%%%%
%%%%%%%%%%%%%%%%%%%%%%%%%%%%%%%%%%%%%%%%%%%%%%%%%%%%%%%%%%%%%%%%%%%%%%%%%%%%%%%


%\bibitem{b2hhhcp}
% N.G.~Deshpande, N.~Sinha and R.~Sinha, \prl{90}{2003}{061802}.
% ==========
% ==========
\bibitem{garmash}
 A.~Garmash {\it et al.} (Belle Collaboration),
 \prd{65}{2002}{092005}.
% ==========
\bibitem{chang}
 K.~Abe {\it et al.} (Belle Collaboration), BELLE-CONF-0317, 2003.
% ==========
\bibitem{garmash2}
 A.~Garmash {\it et al.} (Belle Collaboration), \prd{69}{2004}{012001}.
% ==========
\bibitem{eckhart}
 E. Eckhart {\it et al.} (CLEO Collaboration), \prl{89}{2002}{251801}.
% ==========
\bibitem{aubert}
% B.~Aubert {\it et al.} (BaBar Collaboration), hep-ex/0206004;
 B.~Aubert {\it et al.} (BaBar Collaboration), \prl{91}{2003}{051801}.
% ==========
\bibitem{babar-dalitz}
 B.~Aubert {\it et al.} (BaBar Collaboration), hep-ex/0303022.
% ==========
%\bibitem{KEKB}\
% S.~Kurokawa, \nima{499}{2003}{1}.
% ==========
%\bibitem{Belle}
% A.~Abashian {\it et al.}, \nima{479}{2002}{117}.
% ==========
%\bibitem{GEANT}
% R.~Brun {\it et al.}, GEANT 3.21, CERN Report DD/EE/84-1, 1984.
% ==========
%\bibitem{ArgusF}
% H. Albrecht {\it et al.} (ARGUS Collaboration), \plb{229}{1989}{304}.
% ==========
%\bibitem{CLEO:QQ98}
%  Events are generated with the CLEO group's QQ program \\
% ({\tt http://www.lns.cornell.edu/public/CLEO/soft/QQ}).
% ==========
%\bibitem{PDG}
% K.~Hagiwara {\it et al.} (Particle Data Group), \prd{66}{2002}{10001} and
% 2003 off-year partial update for the 2004 edition (http://pdg.lbl.gov).
% ==========
%\bibitem{dalitz}{R.H.~Dalitz,  Phil. Mag. {\bf 44}, 1068 (1953).}
% ==========
\bibitem{Belle_cont}{Belle-Conf-0410, http://belle.kek.jp\\/conferences/ICHEP2004/}
% ==========
\bibitem{Babar_kpp}{B.~Aubert {\it et al.} (BABAR Collaboration), hep-ex/0408069.}
%\bibitem{d_dp_ana}See for example:
% S.~Kopp {\it et al.} (CLEO Collaboration), \prd{63}{2001}{092001};
% E.M.~Aitala {\it et al.} (E791 Collaboration), \prl{89}{2002}{121801};
% H.~Muramatsu {\it et al.} (CLEO Collaboration), \prl{89}{2002}{251802};
% B.~Aubert {\it et al.} (BABAR Collaboration), hep-ex/0207089.
% ==========
%\bibitem{blatt} J. Blatt and V. Weisskopf, {\it Theoretical Nuclear 
%         Physics}. New York: John Wiley \& Sons (1952).
% ==========
%\bibitem{pilkuhn} H.~Pilkuhn, {\it The Interactions of Hadrons}. 
%         Amsterdam: North-Holland (1967).
% ==========
%\bibitem{kendal}{M.G.~Kendall and A.~Stuart,
%        {\it The Advanced Theory of Statistics}, 2nd ed.
%        (Hafner Publishing, New York, 1968).}
% ==========
%\bibitem{footnote-2} In Dalitz analysis of $D$ meson three-body decays it is
%         common to visualize results of the fit as projections on mass squared
%         variables. Because of the large mass of the $B$ meson, we find
%         that visualization is more convenient in terms of mass variables.
% ==========
%\bibitem{e791-kappa}
% E.M.~Aitala {\it et al.} (E791 Collaboration), \prl{89}{2002}{121801}.
%\bibitem{e791-kappa}E.M.~Aitala {\it et al.} (E791 Collaboration), 
%\prl{89}{2002}{121801}.
% ==========
%\bibitem{beneke-neubert}
% M.~Beneke and M.~Neubert, \npb{675}{2003}{333}.
% ==========
%\bibitem{Flatte}S.M.~Flatt\'e, \plb{63}{1976}{224}.
% ==========
%\bibitem{e791}
% E.M.~Aitala {\it et al.} (E791 Collaboration), \prl{86}{2001}{765}.
% ==========
%\bibitem{phik}
% B.~Aubert {\it et al.} (BaBar Collaboration), \prl{87}{2001}{151801};
% K.-F.~Chen {\it et al.} (Belle Collaboration), \prl{91}{2003}{201801}.
% ==========
\bibitem{b2ps}V.L.~Chernyak, \plb{509}{2001}{273}.
% ==========
\bibitem{maltman}K.~Maltman, \plb{462}{1999}{14}.
% ==========
%\bibitem{b2pt}
% C.S.~Kim, B.H.~Lim and S.~Oh, Eur. Phys. J. {\bf C22}, 683 (2002);\\
% C.S.~Kim, J.P.~Lee and S.~Oh, \prd{67}{2003}{014002}.
% ==========
\bibitem{kskk-cp-belle}
 K.~Abe {\it et al.} (Belle Collaboration), \prl{91}{2003}{261602}.
% ==========
\bibitem{kskk-cp-babar}
 B.~Aubert, {\it et al.} (BaBar Collaboration), \prl{93}{2004}{181805}.
% ==========
%\bibitem{b2pv}
%A. Ali, G. Kramer and C.-D. L\"u,  \prd{58}{1998}{094009};
%Y.H.~Chen, H.Y.~Cheng, B.~Tseng and K.C.~Yang, \prd{60}{1999}{094014};
%M.~Gronau and J.L.~Rosner \prd{61}{2000}{073008};
%H.Y.~Cheng, K.C.~Yang, \prd{62}{2000}{054029};
%D.~Du, H.~Gong, J.~Sun, D.~Yang and G.~Zhu, 
%\prd{65}{2002}{094025}, Erratum-ibid. D{\bf66}, 079904 (2002).
%%


\end{thebibliography}
\end{document}